\documentclass[aps,prd,floatfix,noshowpacs,twocolumn,10pt,showkeys]{revtex4-1}

\usepackage{siunitx}

\usepackage{amsmath}
\usepackage{amssymb}

\usepackage{amsfonts}
\usepackage{mathrsfs}

\usepackage{graphicx}
\usepackage{dcolumn}
\usepackage{bm}
\usepackage{epsfig}
          \usepackage{epstopdf}
\usepackage{enumerate}
 \usepackage{float}
\usepackage{amssymb}

\usepackage{color}

\usepackage[colorlinks=true,linkcolor=darkblue,citecolor=darkblue,urlcolor=darkblue,breaklinks=true]{hyperref}

\usepackage[hyphenbreaks]{breakurl}
\definecolor{darkblue}{rgb}{0,0,0.7}

\def\be{\begin{equation}}
\def\ee{\end{equation}}
\def\bea{\begin{eqnarray}}
\def\eea{\end{eqnarray}}
\def\bml{\begin{flushleft}}
\def\efl{\end{flushleft}}
\def\bmr{\begin{flushright}}
\def\efr{\end{flushright}}
\def\bc{\begin{center}}
\def\ec{\end{center}}
\def\ben{\begin{enumerate}}
\def\een{\end{enumerate}}
\def\bit{\begin{itemize}}
\def\eit{\end{itemize}}

\def\dzn{,\kern-0.1em,}

\def\d#1{{#1\kern-0.4em\char"16\kern-0.1em}}
\def\D#1{{\raise0.2ex\hbox{-}\kern-0.4em 31}}

\def\d{\text{d}}

\newcommand {\apgt} {\ {\raise-.5ex\hbox{$\buildrel>\over\sim$}}\ }
\newcommand {\aplt} {\ {\raise-.5ex\hbox{$\buildrel<\over\sim$}}\ }

\begin{document}

\title{Effects of frustration and Dzyaloshinskii-Moriya interaction on the spin-$1/2$ anisotropic Heisenberg antiferromagnet with the application to $\mbox{La}_2\mbox{CuO}_4$}

\author{Milica Rutonjski}
\affiliation{Department of Physics, Faculty of Sciences, University of Novi Sad, Trg Dositeja
 Obradovi\' ca 4, Novi Sad, Serbia}
\email{milica.rutonjski@df.uns.ac.rs}
\author{Milan Panti\' c}
\affiliation{Department of Physics, Faculty of Sciences, University of Novi Sad, Trg Dositeja
 Obradovi\' ca 4, Novi Sad, Serbia}
\author{Milica Pavkov-Hrvojevi\' c}
\affiliation{Department of Physics, Faculty of Sciences, University of Novi Sad, Trg Dositeja
 Obradovi\' ca 4, Novi Sad, Serbia}






\begin{abstract}
We study the magnetic properties of the two-dimensional anisotropic antiferromagnetic spin-$1/2$ Heisenberg model with Dzyaloshinskii-Moriya interaction and in-plane frustration included. The method of spin Green functions within the framework of Tyablikov's random-phase-approximation decoupling scheme is used in order to derive expressions for the spin-wave spectrum, sublattice magnetization and transition temperature. Based on these expressions we perform a detailed analysis of the influence of varying values of model parameters on its magnetic properties. The model is also applied to the high-$T_{\rm C}$ superconducting parent compound  $\mbox{La}_2\mbox{CuO}_4$ and our results compared to available experimental data.

\end{abstract}

\maketitle








\section{Introduction}
\quad Though the high-$T_{\rm C}$ superconducting parent compound $\mbox{La}_2\mbox{CuO}_4$ has been thoroughly examined in last few decades, the interest in this system, being a good example of the quasi-two-dimensional ($2\rm{D}$) antiferromagnetic Heisenberg model, seems to be inexhaustible. Many models have been proposed in order to describe the magnetic behavior within the $\rm{CuO}_2$ planes, having in mind that it may be associated to the mechanism of the high-$T_{\rm C}$ superconductivity in this compound. Beside the nearest- (NN) and next-nearest-neighbor (NNN) exchange interaction, it is often  suggested that the description without taking into account the Dzyaloshinskii-Moriya (DM) interactions \cite{dzyaloshinsky1958,moriya1960} arising from the spin-orbit coupling may be considered as incomplete \cite{coffey,shekhtman1992,tab1,tab2,silva,parente1,parente2,moskvin2016,moskvin2019}. Namely, in $\mbox{La}_2\mbox{CuO}_4$  due to the spin-orbit couplings arising from the small orthorhombic distortion below the structural tetragonal-orthorhombic transition temperature $T_{\rm tr}=530\,\rm{K}$, $\mbox{CuO}_2$ planes exhibit weak ferromagnetic moment, i. e. all spins cant out of the $\mbox{CuO}_2$ plane by a small angle $\theta$. However, this is not a common feature of the copper oxide layers since in the absence of the orthorhombic distortion DM interaction does not emerge (in $\rm{Sr}_2\rm{CuO}_2\rm{Cl}_2$, for example) \cite{johnston1991,vaknin1997,vaknin1990}. Special interest in DM interaction has also grown due to the possibility of measuring its direction and strength by making use of synchrotron radiation, for the specific class of materials to which $\mbox{La}_2\mbox{CuO}_4$ belongs \cite{dmitrienko2014}. The anisotropy introduced via DM interaction, together with the NN exchange interaction anisotropy \cite{yildirim1994}, is responsible for the existence of the long-range antiferromagnetic order below some nonzero temperature in case of considered $2\rm{D}$ Heisenberg model. Beside these anisotropies, great impact on the magnetic behavior of the studied low-dimensional system has the significant in-plane frustration, which is introduced into the model by taking into account NNN exchange interaction. Namely, though the system is not geometrically frustrated, within the $\mbox{CuO}_2$ planes there exist the conflict of spin orientation due to the antiferromagnetic NNN exchange bonds. The intention to thoroughly analyze the influence of the in-plane frustration on the magnetic properties of the system is corroborated by the fact that frustration is strongly pronounced in two-dimensional quantum magnets \cite{schmidt2017,chubukov1994}, though it is often not taken into account \cite{tab1,tab2,parente1}. Our earlier results \cite{prbspin,prbmag,ssccv,rutonjski2016} also suggest that the model with the NNN interaction gives predictions in better agreement with the experimental data. Beside the aforementioned, the choice of the dominant interactions is supported by the fact that these interactions, especially NN, NNN and DM interaction, but also spin anisotropy to a lesser extent, show a distinctive pressure dependence \cite{pavarini2008,fishman2007}, enabling one to affect the phase transition temperature by applying high pressures on the system. Having this in mind, the purpose of this paper will be to study $2\rm{D}$ antiferromagnetic Heisenberg model with the uniaxial XXZ spin anisotropy, DM interaction and in-plane frustration included. We shall employ the spin Green function method within the framework of Tyablikov's random-phase-approximation (RPA) decoupling scheme. The obtained magnetic properties will be studied in detail in order to describe quantitatively the role of different model parameters. Having at our disposal experimental data for the magnetization temperature dependance in $\mbox{La}_2\mbox{CuO}_4$, we shall apply our model to this compound, as an additional check of our results.

\par The paper is organized as follows: in Sec. II. we present the model Hamiltonian with the dominant exchange interactions and perform the coordinate transformation which facilitates the further calculations. In Sec. III. the main expressions for the quantities to be analyzed are derived by making use of spin Green function method within the RPA scheme. In Sec. IV. the detailed analysis of the numerical results with the emphasis on the comparison of the influence of different system parameters on the studied magnetic properties is given, followed by the application of our results to $\mbox{La}_2\mbox{CuO}_4$ and comparison to available experimental data. The conclusions are briefly stated in Sec. V. 


\section{Model Hamiltonian}

We consider the $S=1/2$ anisotropic Heisenberg antiferromagnet on the rectangular lattice, with the dominant interactions comprised in the following Hamiltonian:  

$$\hat{H}=J\sum_{\bm{n}_a,\bm{\delta}_1}\left(\hat{S}^{x(a)}_{\bm{n}_a}\hat{S}^{x(b)}_{\bm{n}_a+\bm{\delta}_1}+\hat{S}_{\bm{n}_a}^{y(a)}\hat{S}^{y(b)}_{\bm{n}_a+\bm{\delta}_1}\right.+$$ $$\left.+\alpha\,\hat{S}^{z(a)}_{\bm{n}_a}\hat{S}^{z(b)}_{\bm{n}_a+\bm{\delta}_1}\right)-$$
$$-D\sum_{\bm{n}_a,\bm{\delta}_1}\left(\hat{S}_{\bm{n}_a}^{y(a)}\hat{S}_{\bm{n}_a+\bm{\delta}_1}^{z(b)}-\hat{S}_{\bm{n}_a}^{z(a)}\hat{S}_{\bm{n}_a+\bm{\delta}_1}^{y(b)}\right)+$$ 
\be+\frac{J_2}{2}\sum_{\bm{n}_{\nu},\bm{\delta}_2\atop \nu=a,b}\left(\hat{S}_{\bm{n}_{\nu}}^{x(\nu)}\hat{S}^{x(\nu)}_{\bm{n}_{\nu}+\bm{\delta}_2}+\hat{S}_{\bm{n}_{\nu}}^{y(\nu)}\hat{S}^{y(\nu)}_{\bm{n}_{\nu}+\bm{\delta}_2}+\hat{S}_{\bm{n}_{\nu}}^{z(\nu)}\hat{S}^{z(\nu)}_{\bm{n}_{\nu}+\bm{\delta}_2}
\right)\label{hamiltonijan1}\,.\ee

We observe that the Hamiltonian contains the nearest neighbor Dzyaloshinskii-Moriya interaction. The Dzyaloshinskii-Moriya vector $\bm{D}$ is chosen to point along $x$-axis (${\bm{D}}=(D,0,0)$), since its $y$-component would yield the term which does not contribute to the gap, but only has a slight influence on the spin-wave spectrum, wherefore it can be neglected \cite{Korenblit,Entin}. Since the DM interaction itself yields the continuous symmetry in the ground state and therefore precludes the existence of the long AFM order at nonzero temperatures, in 2D models it is common to introduce in Hamiltonian the symmetric pseudodipolar interaction (often denoted by $\tensor{\Gamma}$) which lifts that symmetry. However, one can achieve long-range order also in the case of zero pseudodipolar interaction if the spin anisotropy is included in Hamiltonian  \cite{tab1}, as we do in (\ref{hamiltonijan1}). Namely, the first term in (\ref{hamiltonijan1}) represents the nearest-neighbor anisotropic exchange interaction characterized by the exchange parameter $J$, with the anisotropy parameter denoted by $\alpha$. We take parameter $\alpha$ to be slightly smaller than unity, meaning that the magnetic moments are ordered in the $XY$ plane (easy-plane antiferromagnetism).  Finally, we include the next-nearest-neighbor (NNN) interaction, defined by exchange integral $J_2$. Hereafter we shall define the fundamental energy scale by the nearest-neighbor exchange interaction $J$ and use dimensionless ratios, namely frustration ratio $\lambda=J_2/J$ and DM ratio $d=D/J$ parametrizing the relative strength of DM interaction.  
\par In order to determine the canting angle $\theta$ we shall perform the $180^{\circ}$ rotation of the $b$ sublattice spins around $Z$ axis, which presents the use of extended translational symmetry method of effective lowering the number of sublattices in the system \cite{alistratov}. Minimizing the classical ground state energy, we obtain \be \tan{2\theta}=\frac{2d}{1+\alpha}\,.\label{ugao_teta}\ee It is interesting to notice that the presence of NNN interaction in Hamiltonian does not influence the magnitude of the canting angle.
\par The calculations are easier to be performed if we pass from the initial crystallographic coordinate system to local coordinate systems in which $Z'$ axes coincide with the direction of sublattice magnetization \cite{orbach}. For the sake of simplicity, we take that the spins are antiferromagnetically aligned along the $Y$ axis. 
The transformation matrix for the $a/b$ sublattice reads
\be\left[\begin{array}{c}\hat{S}^{x(a/b)}\\ \hat{S}^{y(a/b)}\\ \hat{S}^{z(a/b)}\end{array}\right]=\left[\begin{array}{ccc}1&0&0\\0&\sin{\theta}&\pm\cos{\theta}\\0&\mp\cos{\theta}&\sin{\theta}\end{array}\right]\left[\begin{array}{c}\hat{\mathcal{S}}^{x(a/b)}\\ \hat{\mathcal{S}}^{y(a/b)}\\ \hat{\mathcal{S}}^{z(a/b)}\end{array}\right]\label{transform_a}\,,\ee
where the upper (lower) sign stands for $a$ $(b)$ sublattice.
\par The Hamiltonian (\ref{hamiltonijan1}) in terms of the new operators $\hat{\mathcal{S}}$ reads
$$\hat{H}/J=\sum_{\bm{n}_a,\bm{\delta}_1}\left[A \left(\hat{\mathcal{S}}_{\bm{n}_a}^{+(a)}\hat{\mathcal{S}}_{\bm{n}_a+\bm{\delta}_1}^{-(b)}+H.c.\right)+\right.$$ $$\left.+C\,\hat{\mathcal{S}}_{\bm{n}_a}^{z(a)}\hat{\mathcal{S}}_{\bm{n}_a+\bm{\delta}_1}^{z(b)}+M\left(\hat{\mathcal{S}}_{\bm{n}_a}^{+(a)}\hat{\mathcal{S}}_{\bm{n}_a+\bm{\delta}_1}^{+(b)}+H.c.\right)\right]+$$
\be+\frac{\lambda}{2}\sum_{\bm{n}_{\nu},\bm{\delta}_2\atop \nu=a,b}\left[\frac12\left(\hat{\mathcal{S}}_{\bm{n}_{\nu}}^{+(\nu)}\hat{\mathcal{S}}^{-(\nu)}_{\bm{n}_{\nu}+\bm{\delta}_2}+H.c.\right)+\hat{\mathcal{S}}_{\bm{n}_{\nu}}^{z(\nu)}\hat{\mathcal{S}}^{z(\nu)}_{\bm{n}_{\nu}+\bm{\delta}_2}
\right]\,,\label{hamiltonijan2}\ee
where we use the following notation
\begin{equation}
\begin{aligned} A&=\frac12-\frac{(1+\alpha)\cos^2{\theta}}{4}-d\,\frac{\sin{2\theta}}{4}\,,\\
 C&=\alpha-(1+\alpha)\cos^2{\theta}-d\,\sin{2\theta}\,,\\
 M&=\frac{(1+\alpha)\cos^2{\theta}}{4}+d\,\frac{\sin{2\theta}}{4}\,.\label{oznakeACM}\end{aligned}
\end{equation}
By making use of the transformation (\ref{transform_a}) we have effectively described the system by the easy-axis model. Transformed Hamiltonian (\ref{hamiltonijan2}) will present the starting point in our calculations.


\section{Spin-wave spectrum, magnetization and related quantities}

\par In order to derive spin-wave spectrum, we use the spin Green function method within the framework of Tyablikov's decoupling approximation \cite{tjablikov}. Due to the structure of Hamiltonian, we obtain the following system of four equations of motion for Green functions $\langle\langle\hat{\mathcal{S}}^{\pm(a/b)}|\hat{B}\rangle\rangle_{\bm{k},E}\equiv G^{\pm}_{a/b}$(explicit dependence on wave-vector and energy will be hereafter omitted for brevity):
\begin{equation}
\begin{aligned}
(\omega+D_{\bm{k}})\,G^+_a-P_{\bm{k}}\,G^+_b-R_{\bm{k}}\,G^-_b & =  \frac{i}{2\pi}\langle[\hat{\mathcal{S}}^{+(a)},\hat{B}]\rangle \\
(\omega-D_{\bm{k}})\,G^-_a+R_{\bm{k}}\,G^+_b+P_{\bm{k}}\,G^-_b & =  \frac{i}{2\pi}\langle[\hat{\mathcal{S}}^{-(a)},\hat{B}]\rangle  \\
-P_{\bm{k}}\,G^+_a-R_{\bm{k}}\,G_-^a+(\omega+D_{\bm{k}})\,G^+_b & = \frac{i}{2\pi}\langle[\hat{\mathcal{S}}^{+(b)},\hat{B}]\rangle \\
 R_{\bm{k}}\,G^+_a+P_{\bm{k}}\,G_-^a+(\omega-D_{\bm{k}})\,G^-_b & =  \frac{i}{2\pi}\langle[\hat{\mathcal{S}}^{-(b)},\hat{B}]\rangle  \label{sistemGF}
\end{aligned}
\end{equation}

where
\begin{equation}
\begin{aligned} D_{\bm{k}}&=J\left[z_1 C+z_2 \lambda (1-\gamma_{2\bm{k}})\right]\sigma\,,\\
 P_{\bm{k}}&=2 z_1 JA \gamma_{1\bm{k}}\sigma\,,\\
 R_{\bm{k}}&=2 z_1 JM \gamma_{1\bm{k}}\sigma\,.\label{funkcijeDPR}\end{aligned}
\end{equation}
In Eqs. (\ref{funkcijeDPR}) $\sigma=\langle\hat{\sigma}^z\rangle$ denotes the sublattice magnetization (in the absence of external magnetic field $\sigma^{(a)}=\sigma^{(b)}=\sigma$), $z_{1/2}=4$ defines the number of nearest and next-nearest neighbor sites, while quantities $\gamma_{1/2\,{\bm{k}}}$ present geometric factors given by
\begin{equation}\begin{aligned}\gamma_{1\bm{k}}&=\cos{\frac{k_xa_0}{2}}\cos{\frac{k_yb_0}{2}}\,,\\ \gamma_{2\bm{k}}&=\frac12(\cos{k_xa_0}+\cos{k_yb_0})\,,\label{gama12}\end{aligned}\end{equation}
where $a_0$ and $b_0$ denote the lattice parameters.

Spin-wave spectrum obtained from the system of equations (\ref{sistemGF}) for the in-plane mode is given by
\be \omega_{1\bm{k}}=z_1\sigma J\sqrt{[C+\lambda (1-\gamma_{2\bm{k}})-2\gamma_{1\bm{k}}A]^2-4\gamma_{1\bm{k}}^2M^2}\,,\label{inplane_mode}\ee
while for the out-of-plane mode reads
\be\omega_{2\bm{k}}=z_1\sigma J\sqrt{[C+\lambda (1-\gamma_{2\bm{k}})+2\gamma_{1\bm{k}}A]^2-4\gamma_{1\bm{k}}^2M^2}\,.\label{outofplane_mode}\ee
The corresponding zone-center ($\bm{k}=0$) energy gaps are
\be\omega_{1/2\,\bm{k}=0}=z_1\sigma J\sqrt{(C\mp2A)^2-4M^2}\,.\label{gepovi}\ee
In order to determine sublattice magnetization from the well-known formula
\be\sigma=\frac12-\frac1N\sum_{\bm{k}}\langle\hat{\mathcal{S}}^-\hat{\mathcal{S}}^+\rangle_{\bm{k}}\,,\,\label{osnovna_mag}\ee we calculate the Green function $\langle\langle\hat{\mathcal{S}}^{+(a)}|\hat{\mathcal{S}}^{-(a)}\rangle\rangle_{\bm{k},E}\equiv G^{+-}_{aa}$ from the system (\ref{sistemGF}) taking for the operator $\hat{B}\equiv\hat{\mathcal{S}}^{-(a)}$.
Using the standard procedure to obtain the correlation function $\langle\hat{\mathcal{S}}^-\hat{\mathcal{S}}^+\rangle_{\bm{k}}$, we derive the following expression for the sublattice magnetization
$$\sigma=\left[\frac{1}{N}\sum_{\bm{k}}\left(\frac{\mathcal{A}_{\bm{k}}-\mathcal{C}_{\bm{k}}}{\sqrt{(\mathcal{C}_{\bm{k}}-\mathcal{A}_{\bm{k}})^2-\mathcal{M}_{\bm{k}}^2}}\coth\frac{\beta\omega_{1\bm{k}}}{2}-\right.\right.$$ \be\left.\left.-\frac{\mathcal{A}_{\bm{k}}+\mathcal{C}_{\bm{k}}}{\sqrt{(\mathcal{C}_{\bm{k}}+\mathcal{A}_{\bm{k}})^2-\mathcal{M}_{\bm{k}}^2}}\coth\frac{\beta\omega_{2\bm{k}}}{2}\right)\right]^{-1}\,,\label{magnetizacija}\ee
where \begin{equation}\begin{aligned}\mathcal{A}_{\bm{k}}&=2JA\gamma_{1\bm{k}}\,,\\ \mathcal{C}_{\bm{k}}&=J[C+\lambda (1-\gamma_{2\bm{k}})]\,,\\ \mathcal{M}_{\bm{k}}&=2JM\gamma_{1\bm{k}}\label{kaligrafskeoznake}\end{aligned}\end{equation} and $\beta=1/k_{\rm B}T$.
The temperature at which magnetization vanishes is obtained from (\ref{osnovna_mag}), using the expansion $\coth x\sim1/x$ valid at high temperatures, wherefrom
\be\theta_N=\left[\frac{1}{2N}\sum_{\bm{k}}\left(\frac{\mathcal{A}_{\bm{k}}-\mathcal{C}_{\bm{k}}}{(\mathcal{C}_{\bm{k}}-\mathcal{A}_{\bm{k}})^2-\mathcal{M}_{\bm{k}}^2}-\frac{\mathcal{A}_{\bm{k}}+\mathcal{C}_{\bm{k}}}{(\mathcal{C}_{\bm{k}}+\mathcal{A}_{\bm{k}})^2-\mathcal{M}_{\bm{k}}^2}\right)\right]^{-1}\,.\label{Nelova_temp}\ee
We shall now analyze these results.


\section{Analysis of results}

\par
\begin{figure} 
\includegraphics[scale = 0.8]{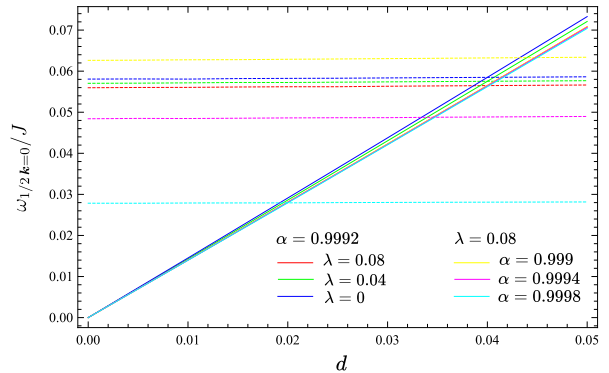}
{\caption{\label{Fig1} In-plane (dashed lines) and out-of-plane (solid lines) reduced zone-center energy gaps $\omega_{{1/2}\,{\bm{k}}=0}/J$ at $T=0\,{\rm{K}}$ vs. parameter $d$.}}
\end{figure}

\begin{figure} 
\includegraphics[scale = 0.8]{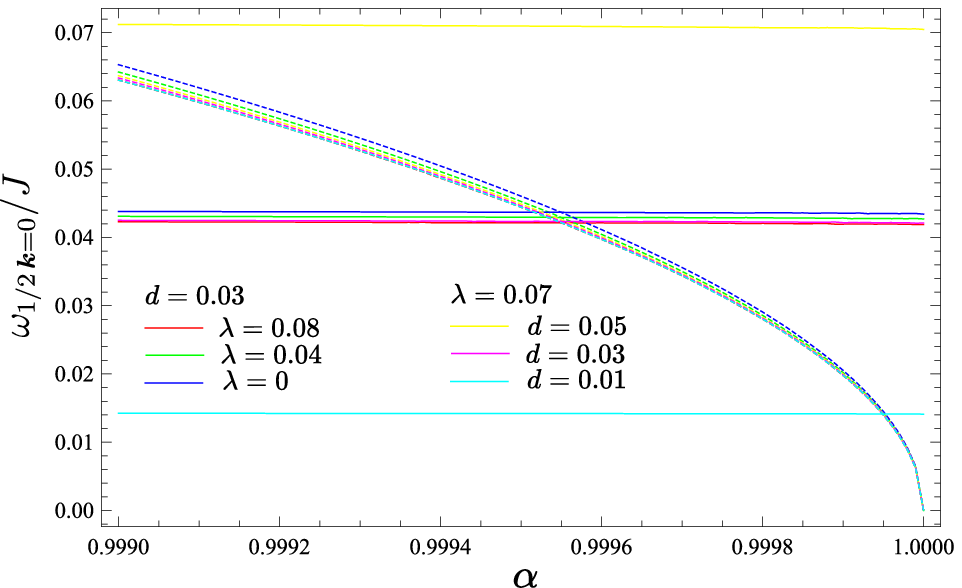}
{\caption{\label{Fig2} In-plane (dashed lines) and out-of-plane (solid lines) reduced zone-center energy gaps $\omega_{{1/2}\,{\bm{k}}=0}/J$ at $T=0\,{\rm{K}}$ vs. parameter $\alpha$.}}
\end{figure}

\begin{figure} 
\includegraphics[scale = 0.8]{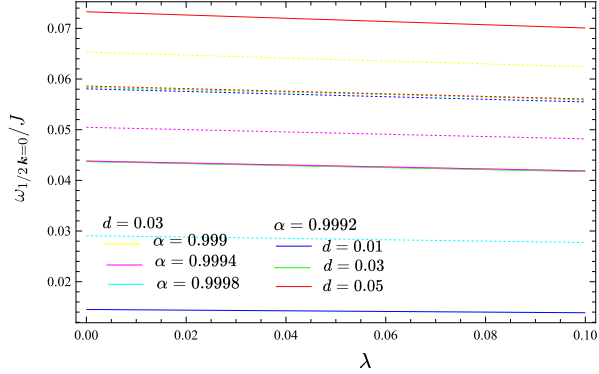}
{\caption{\label{Fig3} In-plane (dashed lines) and out-of-plane (solid lines) reduced zone-center energy gaps $\omega_{{1/2}\,{\bm{k}}=0}/J$ at $T=0\,{\rm{K}}$ vs. parameter $\lambda$.}}
\end{figure}

 We shall start the analysis with the numerical calculation of the reduced zero-temperature energy gaps in the long-wavelength limit $\omega_{1/2\,\bm{k}=0}/J$ as functions of the introduced model parameters. The range of parameter values is chosen having in mind the parameter sets based on the experimental measurements describing the real magnetic systems ($\mbox{La}_2\mbox{CuO}_4$ for example, as will be evident later). In Fig. \ref{Fig1} we present the energy gaps dependance on DM parameter $d$, for different values of the other two parameters $\alpha$ and $\lambda$. By inspection of Fig. \ref{Fig1} we infer that for the non-zero values of $d$ both spin-wave modes possess gap, while when $d$ vanishes, in-plane mode (almost independent on $d$) possesses gap of magnitude $\omega_{1\,\bm{k}=0}\propto z_1\sigma_0\sqrt{1-\alpha}$ and out-of-plane mode (almost linear in $d$) becomes gapless, i.e. Goldstone mode appears. These observations agree with the ones from \cite{tab1}, suggesting that the NNN interaction does not change qualitatively the energy gap behavior. However, in order to describe quantitatively the influence of the frustration parameter $\lambda$ we here perform the more detailed analysis by comparison of the influence of the parameters $\lambda$ and $\alpha$ on the studied dependance. As can be seen, the influence of the NNN interaction parameter $\lambda$ on the nonvanishing gap is much smaller than the influence of the anisotropy parameter $\alpha$. Similar observations  can be made based on the energy gaps dependance on anisotropy parameter $\alpha$, presented in Fig. \ref{Fig2}. Contrary to previous case, the in-plane mode is the one which in case of isotropic model ($\alpha=1$) becomes gapless, while the out-of-plane gap remains finite $\omega_{2\,\bm{k}=0}\propto z_1\sigma_0\sqrt{d\cos{3\theta}+(d^2+(d^2-1)\cos{2\theta})\sin{\theta}}$. The influence of parameter $\lambda$ on the nonvanishing gap is less prominent than the influence of DM parameter. Finally, we consider the energy gaps dependance on frustration parameter $\lambda$ (Fig. \ref{Fig3}). In the considered range of $\lambda$ both energy gaps are finite and almost independent on that parameter.

\bc
\ec

\bc
\ec

\bc
\ec

\begin{figure} 
\includegraphics[scale = 0.8]{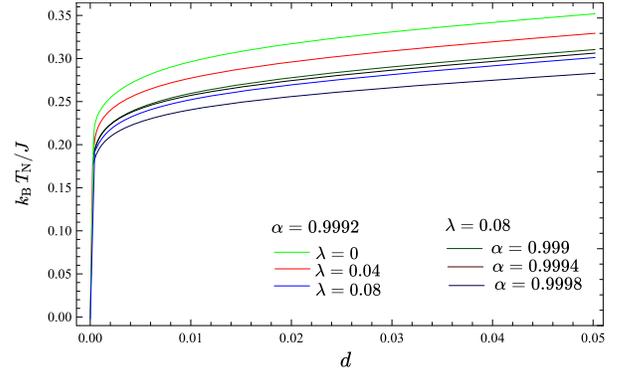}
{\caption{\label{Fig4} Reduced transition temperature vs. parameter $d$ for for different values of parameters $\alpha$ and $\lambda$.  }}
\end{figure}

\begin{figure} 
\includegraphics[scale = 0.8]{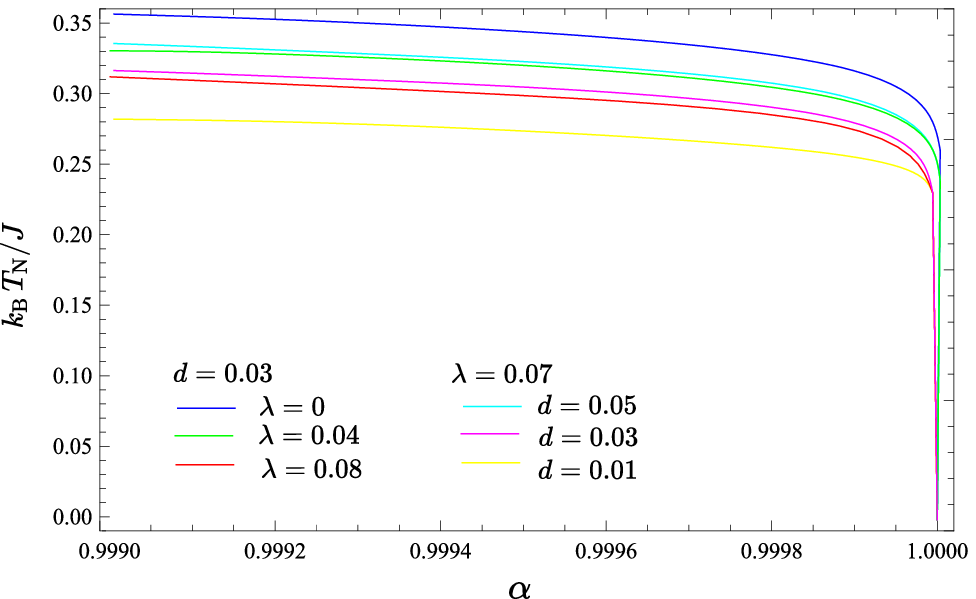}
{\caption{\label{Fig5} Reduced transition temperature vs. parameter $\alpha$ for for different values of parameters $d$ and $\lambda$. }}
\end{figure}

\begin{figure} 
\includegraphics[scale = 0.8]{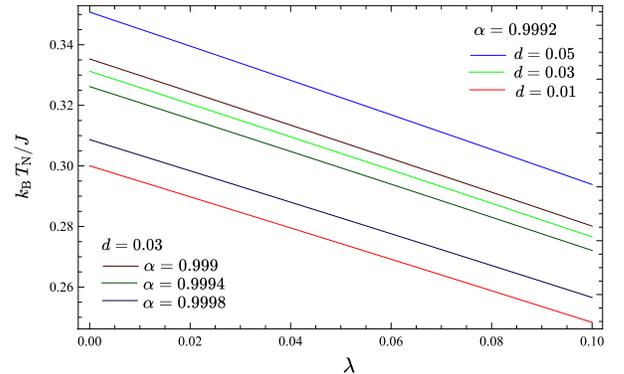}
{\caption{\label{Fig6}Reduced transition temperature vs. parameter $\lambda$ for for different values of parameters $\alpha$ and $d$.  }}
\end{figure}

It is further interesting to examine the dependence of transition temperature on various parameters of the system. In Figs. \ref{Fig4}-\ref{Fig6} we present the reduced N\' eel temperature dependence on parameters $d$, $\alpha$ and $\lambda$ respectively, as a function of several values of other two parameters. Fig. \ref{Fig4} shows that with the increase of DM parameter $d$ the transition temperature also increases. Due to the fact that Goldstone mode in spin-wave spectrum appears when $d$ vanishes, transition temperature drops to zero in that limit. From Fig. \ref{Fig5} we can see that the decrease of parameter $\alpha$ (i. e. the increase of spin anisotropy) increases the transition temperature. On the other hand, $T_N$ becomes zero for $\alpha=1$. From Fig. \ref{Fig6} we see that the increase of parameter $\lambda$ yields the decrease of the transition temperature, since the in-plane frustration disorders the system. The transition temperature is however not suppressed to zero, since in the considered parameter range energy gaps remain finite. 

\bc
\ec

\bc
\ec

\bc
\ec

\begin{figure} 
\includegraphics[scale = 0.8]{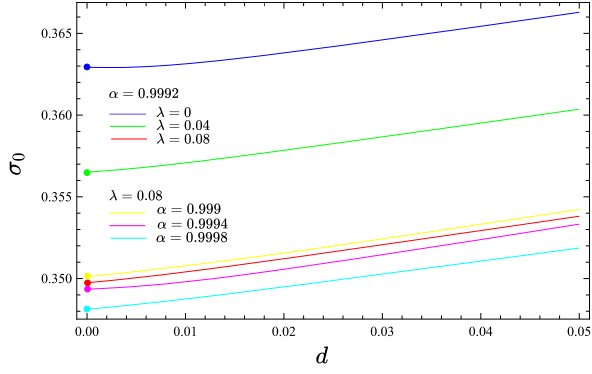}
{\caption{\label{Fig7} Sublattice magnetization $\sigma$ at $T=0\,\rm{K}$ vs. parameter $d$ for different values of parameters $\alpha$ and $\lambda$. Dots emphasize the nonzero values of $\sigma_0$ for $d=0$.}}
\end{figure}

\begin{figure} 
\includegraphics[scale = 0.8]{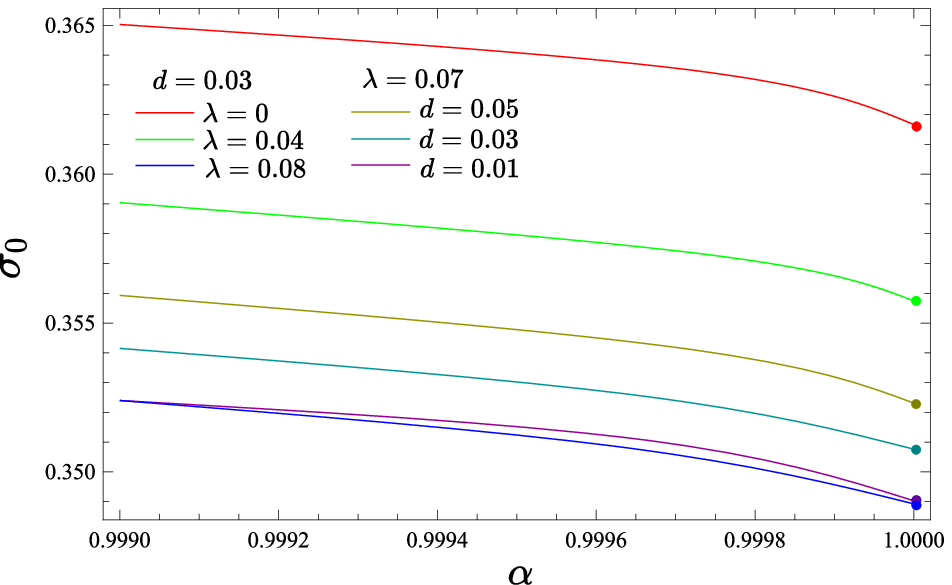}
{\caption{\label{Fig8} Sublattice magnetization $\sigma$ at $T=0\,\rm{K}$ vs. parameter $\alpha$ for different values of parameters $d$ and $\lambda$.  Dots emphasize the nonzero values of $\sigma_0$ for $\alpha=1$.}}
\end{figure}

\begin{figure} 
\includegraphics[scale = 0.8]{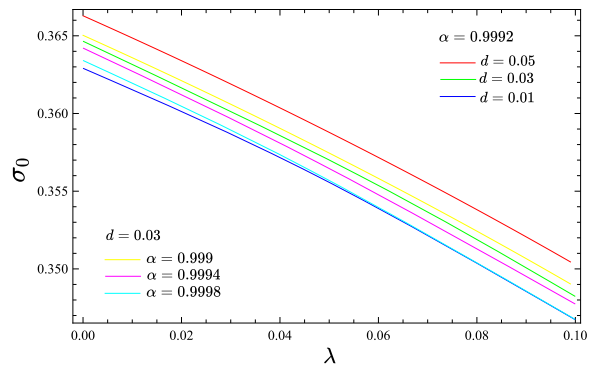}
{\caption{\label{Fig9} Sublattice magnetization $\sigma$ at $T=0\,\rm{K}$ vs. parameter $\lambda$ for different values of parameters $d$ and $\alpha$. }}
\end{figure}

We now calculate zero-temperature antiferromagnetic order parameter $\sigma_0$, by making use of the expression (\ref{magnetizacija}). The behavior of $\sigma_0$ as a function of parameter $d$ for several value of parameters $\alpha$ and $\lambda$ is presented in Fig. \ref{Fig7}. Analogous dependencies of the zero-temperature magnetization on parameters $\alpha$ and $\lambda$ for different values of two other Hamiltonian parameters are shown in Figs. \ref{Fig8} and \ref{Fig9}. By inspection of Figs. \ref{Fig7}-\ref{Fig9} one can see that the increase of DM parameter tends to increase $\sigma_0$, while the increase of either the frustration parameter $\lambda$ or parameter $\alpha$ tend to decrease the order parameter. It should be noted that though the transition temperature of the isotropic model ($\alpha=1$) and the model without DM interaction ($d=0$) equals zero, the sublattice magnetization at $T=0\,\rm{K}$ is finite in both cases, i.e. the long-range order exists at zero temperature but is destroyed by the thermal fluctuations for nonzero temperatures. 
Finally, the analysis of Figs. \ref{Fig7} and \ref{Fig8} shows that the influence of the frustration parameter $\lambda$ on the zero-temperature magnetization is more pronounced than in the case of energy gaps.

\bc
\ec

\bc
\ec

\bc
\ec

\begin{figure} 
\includegraphics[scale = 0.8]{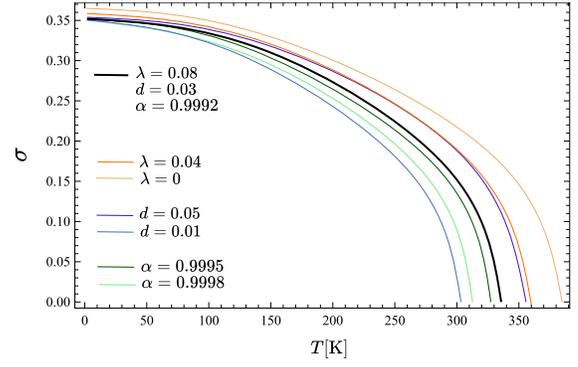}
{\caption{\label{Fig10} Sublattice magnetization $\sigma$ vs. temperature $T$, for the NN interaction $J=100\,\rm{meV}$. The lines in color are denoted by the value of the parameter by which they differ from the black line.}}
\end{figure}

The influence of the frustration, DM interaction and spin anisotropy on the sublattice magnetization in the wide temperature range $0\leq T\leq T_{\rm{N}}$ is shown in Fig. \ref{Fig10}. Due to the self-consistency of the Eq. (\ref{magnetizacija}), the iterative procedure had to be applied. The value of nearest-neighbor exchange interaction $J$ is chosen to be $100\,\rm{meV}$. 

\bc
\ec

\subsection{Application to $\rm{La}_2\rm{CuO}_4$}

As it was stated before, the magnetic properties of the high-$T_{\rm C}$ superconducting parent compound $\mbox{La}_2\mbox{CuO}_4$ can be considered as quasi-two-dimensional system \cite{tab1,tab2} described by the Hamiltonian (\ref{hamiltonijan1}).
In order to determine model parameters, we use experimental data for the gaps of the in-plane and out-of-plane polarized magnons at $100\,K$ \cite{BirgenauZeitschrift}, with magnitudes $\omega_{1\,{\bm{k}}=0}=2.3\,{\mbox{meV}}$ and $\omega_{2\,{\bm{k}}=0}=5\,{\mbox {meV}}$ respectively. Using self-consistent procedure to determine $J$ and $\alpha$, and choosing other two parameters to reproduce correct value for the N\' eel temperature, we obtain: \begin{equation} \begin{aligned} J&=110.251\,\rm{meV}\,\,\,\,\,\alpha=0.99987\\
d&=0.035\,\,\,\,\,\,\,\,\,\,\,\,\,\,\,\,\,\,\,\,\,\,\,\,\,\lambda=0.09 \label{parametri2jednacine}\end{aligned}\end{equation} 
wherefrom the canting angle in $\mbox{La}_2\mbox{CuO}_4$ reads $\theta \approx 0.017^{\circ}$. While the parameters $\alpha$ and $d$ are of the same order of magnitude as those proposed in \cite{tab1}, the superexchange value $J$ is lower than the one quoted in \cite{tab1}, the latter leading to the significant overestimation (over $30\%$) of the transition temperature.

\begin{figure} 
\includegraphics[scale = 0.65]{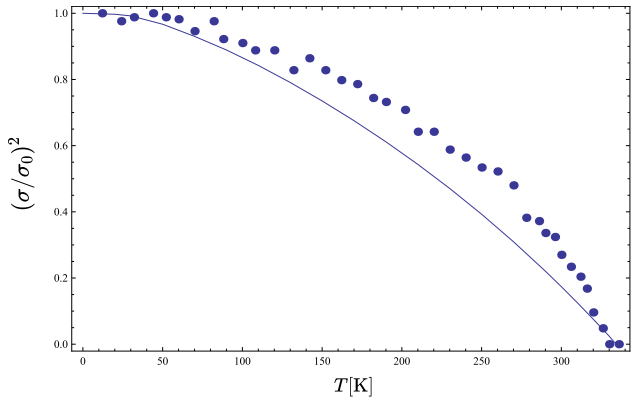}
{\caption{\label{Fig11} Square of the relative magnetization $(\sigma/\sigma_0)^2$ vs. temperature $T$. Full circles denote the experimental data from \cite{keimermagnetizacija}. Solid line represents the theoretical result based on expression (\ref{magnetizacija}).}}
\end{figure}

\begin{figure} 
\includegraphics[scale = 0.8]{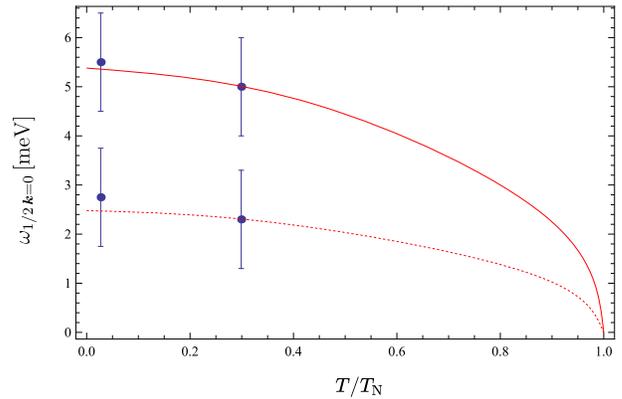}
{\caption{\label{Fig12} In-plane (dashed curve) and out-of-plane (full curve) zone-center energy gaps $\omega_{{1/2}\,{\bm{k}}=0}$ vs. relative temperature $T/T_{\rm N}$. Full circles denote the experimental data from \cite{BirgenauZeitschrift}. Solid lines represent the theoretical result based on expression (\ref{gepovi}).}}
\end{figure}

An important check of our results is the comparison to the magnetization temperature dependence obtained by neutron scattering reported in \cite{keimermagnetizacija}. As can be seen from Fig. \ref{Fig11}, our model compares favorably to the experiment. Finally, we present the temperature dependence of the energy gaps of both modes in excitation spectrum (Fig. \ref{Fig12}). It can be seen that our results are in agreement with the experimental data given in \cite{BirgenauZeitschrift}. 

\bc
\ec

\bc
\ec

\section{Conclusions}

We study the two-dimensional anisotropic Heisenberg antiferromagnet, with the Dzyaloshinskii-Moriya interaction and in-plane frustration included. By making use of spin Green function method within the Tyablikov's decoupling approximation, we obtain the expressions for the spin-wave energy gaps, sublattice magnetization and transition temperature. Detailed comparison of the influence of model parameters on the magnetic properties of the system is performed. We conclude that these parameters have the opposite impact on the long-range antiferromagnetic order, whereby the increase of the frustration parameter $\lambda$ and spin anisotropy parameter $\alpha$ destabilize the system, while the increase of DM parameter $d$ stabilizes the system.

We also apply our results to the high-$T_{\rm{C}}$ superconducting parent compound $\mbox{La}_2\mbox{CuO}_4$, a layered copper oxide which has been both theoretically and experimentally investigated in last decades with undiminished interest. The magnetization curve and energy gaps temperature dependance based on our calculations agree favorably with the experimental data. However, new experimental results, such as measurements of the strength and direction of DM interaction in $\mbox{La}_2\mbox{CuO}_4$ as well as high pressure induced variation of the in-plane frustration, would serve as a further check of our results.


\section*{Acknowledgments}
We thank Dr. Slobodan Rado\v sevi\'c for helpful discussions. This work was supported by the Serbian Ministry of
Education, Science and Technological Development under Contract No. 451-03-68/2020-14/ 200125.

\bibliography{Refs}

\end{document}